# PERIODONTITIS AND PREECLAMPSIA IN PREGNANCY: A SYSTEMATIC REVIEW AND META-ANALYSIS

**Quynh-Anh LE[1], Rahena AKHTER[2], Kimberly Mathieu COULTON[2], Ngoc Truong Nhu VO[3], Le Thi Yen DUONG[4], Hoang Viet NONG[4], Albert YAACOUB[5], George CONDOUS[6], Joerg EBERHARD[7], Ralph NANAN[8]**

[1] PhD candidate, DDS, School of Dentistry and the Charles Perkins Center, Faculty of Medicine and Health, The University of Sydney, Sydney, New South Wales, Australia.

[2] PhD, School of Dentistry and the Charles Perkins Center, Faculty of Medicine and Health, The University of Sydney, Sydney, New South Wales, Australia.

[3] Associate Professor, Department of Pediatric Dentistry, School of Odonto-Stomatology, Hanoi Medical University, Hanoi, Vietnam.

[4] DDS, School of Odonto-Stomatology, Hanoi Medical University, Hanoi, Vietnam.

[5] MPH, Nepean Centre for Oral Health, Nepean hospital, Kingswood, New South Wales, Australia.

[6] Associate Professor, Acute Gynaecology, Early Pregnancy, and Advanced Endoscopy Surgery Unit, Nepean Hospital, Sydney Medical School Nepean, University of Sydney, Sydney, Australia.

21    [7] Professor, School of Dentistry and the Charles Perkins Center, Faculty of Medicine and

22    Health, The University of Sydney, Sydney, New South Wales, Australia.

23    [8] Professor, Sydney Medical School Nepean and Charles Perkins Center Nepean, The

24    University of Sydney, Sydney, New South Wales, Australia.

**CONFLICT OF INTEREST**

32    The authors report no conflict of interest.

* Corresponding author    Dr Quynh-Anh Le

Address                   School of Dentistry, Faculty of Medicine and Health, The

                          University of Sydney, Sydney, New South Wales, Australia.

Telephone                 +61415076584

Email address             Qule7436@uni.sydney.edu.au

38    Total word count of abstract: 231

39    Total word count of main text: 6192

40 **ABSTRACT**

41 **Objectives:** A conflicting body of evidence suggests localized periodontal inflammation to

42 spread systemically during pregnancy inducing adverse pregnancy outcomes. This systematic

43 review and meta-analysis aims to specifically evaluate the relationship between periodontitis

44 and preeclampsia.

45 **Data sources:** Electronic searches were carried out in Medline, Pubmed, Embase, LiLacs,

46 Cochrane Controlled Clinical Trial Register, CINAHL, ClinicalTrials.gov and Google

47 Scholar with no restrictions on year of publication.

48 **Study eligibility criteria**: We identified and selected observational case-control and cohort

49 studies that analyzed the association between periodontal disease and preeclampsia.

50 **Study appraisal and synthesis methods:** This meta-analysis was conducted according to

51 PRISMA guidelines and MOOSE checklist. Pooled odds ratios, mean difference, and 95%

52 confidence intervals were calculated using the random effect model. Heterogeneity was

53 tested with Cochran's Q statistic.

54 **Results:** Thirty studies including six cohort and twenty-four case-control studies were

55 selected. Periodontitis was significantly associated with increased risk for preeclampsia (OR

56 3.18, 95% CI 2.26 – 4.48, p<0.00001), especially in a subgroup analysis including cohort

57 studies (OR 4.19, 95% CI 2.23 – 7.87, p < 0.00001). The association was even stronger in a

58 subgroup analysis with lower middle-income countries (OR 6.70, 95% CI 2.61 – 17.19, p <

59 0.0001).

60 **Conclusion:** Periodontitis appears as a significant risk factor for preeclampsia, which might

61 be even more pronounced in lower middle-income countries. Future studies to investigate if

62 maternal amelioration of  periodontitis prevents preeclampsia might be warranted.

63 **Keywords:** Periodontitis, periodontal disease, preeclampsia, pre-eclampsia, hypertension,

64 pregnancy outcome.

65 **Abbreviations and Acronyms:**

66 PD:    pocket depth

67 CAL:   clinical attachment loss

68

69 **INTRODUCTION**

70 Preeclampsia is the onset of pregnancy-related hypertensive disorder and proteinuria arising

71 most commonly after 20 weeks of gestation, which could lead to eclampsia and induce

72 maternal and perinatal morbidity and mortality. The prevalence of preeclampsia is between

73 2% to 8% of all pregnancies worldwide.[1] Preeclampsia affected pregnancies had a higher risk

74 of poor maternal outcomes including cerebrovascular bleeding, HELLP syndrome,

75 eclampsia, poorer outcomes of their offspring including premature birth, intrauterine growth

76 restriction and the complications may manifest over years postpartum.[2, 3]

77

78 Contributing to USD 6.4 billion short-term estimated costs for preeclamptic pregnancies in

79 US healthcare system, USD1.03 billion were spent on maternal healthcare and USD1.15

80 billion were expended for infants born to these mothers while the remaining expenses were

81 for peripartum and postpartum care[4].

82

83 Herein, it is of utmost importance to manage risk factors of preeclampsia to improve maternal

84 and perinatal outcomes as well as lessen the burden on the health economic aspects.

85

86 Depending on geographical regions approximately 14.2% and 54.8% of pregnant women

87 suffer from periodontal disease. [5-7]Especially, periodontitis, a more severe type of periodontal

88 diseases affecting 11% of the pregnant women, can cause the destruction of periodontal

89 tissue and cause systemic dissemination of bacteria and other inflammatory mediators. [8, 9]

90 Systemic inflammatory processes triggered by focal periodontal infections have been

91 attributed to cardiovascular, cerebrovascular diseases and respiratory diseases.[10] Periodontitis

92 has independently been linked to several pregnancy complications such as preterm birth, low

93 birth weight and gestational diabetes.[11, 12]

94

95 Socioeconomic status is a recognized factor associated with medical outcomes, including

96 pregnancy outcomes.[13] Women with lower socioeconomic status are at a higher risk of

97 pregnancy complications such as gestational diabetes, preterm delivery and preeclampsia.[14-16]

98 Women with high socioeconomic status have a statistically significant reduced risk of

99 preeclampsia with an odds ratio of 0.899 (95% CI, 0.862 – 0.937, $p < 0.001$) compared to

100 women with lower socioeconomic statuses.[17] At the same time, the proportion of

101 periodontitis in pregnancy is linked to low socioeconomic status with 42.6%, compared to

102 high socioeconomic status with 15.0%.

103

104 Two previous meta-analyses both published in 2013 reported positive associations between

105 preeclampsia and periodontitis with OR 2.17, 95% CI 1.38-3.41, p=0008 and OR of 2.79,

106 95% CI 2.01-3.01, p<0.0001, but did not consider socioeconomic factors.[18,19] Since then

107 further case-control and cohort studies have been published on this research topic. [20,21,22]

108 Nonetheless, the causal relationship between periodontal disease and preeclampsia remains

109 unclear [23,24]In this review, we included all available new studies, to re-evaluate the potential

110 association between periodontitis and preeclampsia and also to take socioeconomic factors

111 into consideration. [25]

112

113

114 **METHODS**

115

116 **Eligibility criteria**

117 The studies were screened according to the following inclusion criteria: (1) case-control,

118 prospective cohort study, (2) studies analysing the association between periodontal disease

119    and preeclampsia, (3) study population was pregnant women without systemic diseases, and

120    (4) data was presented in such a way where Odds Ratio and 95% Confidence Interval could

121    be calculated. Studies were excluded if they did not report adequate data or outcome of

122    interest.

123    **Information sources**

124    We followed the Preferred Reporting Items for Systematic Reviews and Meta-Analysis

125    (PRISMA) guidelines with the checklist of 27 items to conduct our study.[26] We also adopted

126    the MOOSE checklist for Meta-analysis of Observational studies.[27] A systematic search of

127    the electronic database including Medline (from 1950), Pubmed (from 1946), Embase (from

128    1949), Lilacs, Cochrane Controlled Clinical Trial Register, CINAHL, ClinicalTrials.gov and

129    Google Scholar (from 1990) to identify relevant articles .

130    **Search strategy**

131    We used the following search terms:  periodontitis, periodontal disease, preeclampsia, pre-

132    eclampsia, pregnancy outcomes, pregnancy complications, hypertension. The combinations

133    of search terms were used to explore above databases. The search strategy was peer reviewed

134    by two independent reviewers (QA and LD). The reference lists of relevant articles were also

135    scanned for appropriate studies. No language restrictions were adopted in either the search or

136    study selection. No search for unpublished literature was carried out. Authors were contacted

137    for translation and information.

138

139    **Study selection**

140    Two independent reviewers (LD and HN) reviewed the titles, abstracts and methods of

141    retrieved results to assess for the eligibility criteria. When there was a disagreement in a

142    selection process between reviewers, consensus with the third reviewer (QA) was obtained.

143

**Data extraction**

Data extraction was carried out using a standardized extraction form, collecting information on the first author's name, publication year, study design, number of cases, number of controls, total sample size, country, national income (according to World Bank classification[28]), mean age, the risk of estimates or data used to calculate the risk estimates, Cis or data used to generate CI. The researchers cross-checked all extracted data and discussed if there were disagreements.

**Assessment of risk of bias**

Risk of bias was executed using the Newcastle Ottawa Scale by two reviewers (QA and LD)[29] with disagreements resolved by consensus attainment between reviewers. This Scale has three components including Selection, Comparability and Outcome/Exposure assessment with maximum overall score of nine. Studies were rated as low risk of bias if they received nine score, moderate risk of bias if they received seven or eight score and high risk of bias if they received less than seven scores.

**Data synthesis**

Data were imported in a statistical software (RevMan, Version 5, 2008, The Nordic Cochrane Center, The Cochrane Collaboration, Copenhagen, Denmark). Pooled Odds Ratios, mean difference, and 95% Confidence Intervals were calculated for the association between periodontitis and preeclampsia using a random effects model. The pooled effect was considered significant if p-value was less than 0.05. Forest plots for primary analysis and subgroup analysis show the raw data, Odds Ratio and CIs, Means and SDs for the chosen effect, heterogeneity statistic ($I^2$), total number of participants per group, overall Odds Ratio and Mean difference.

169     Subgroup analysis was carried out according to the study design (case-control or cohort),

170     definition of periodontitis (defined by pocket depth (PD) and/or clinical attachment loss [30]),

171     mean CAL, mean PD, national income (high-income or middle-income or low-income

172     countries).

173     Heterogeneity was tested with Cochran's Q statistic, with P<0.10 indicating heterogeneity,

174     and quantified the degree of heterogeneity using the $I^2$ statistic, which represents the

175     percentage of the total variability across studies which is due to heterogeneity. $I^2$ values of

176     25, 50 and 75% corresponded to low, moderate and high degrees of heterogeneity

177     respectively[31]. We quantified publication bias using the Egger's regression model with the

178     effect of bias assessed using the fail-safe number method.[32] The fail-safe number was the

179     number of studies that we would need to have missed for our observed result to be nullified

180     to statistical non-significance at the p<0.05 level. Publication bias is generally regarded as a

181     concern if the fail-safe number is less than 5n+10, with n being the number of studies

182     included in the meta-analysis.[33] Publication bias was assessed using Stata (16.1, StataCorp

183     LLC, College Station, TX).

184

185     **RESULTS**

186

187     **Study selection**

188     A total of 3450 articles were found through the manual and electronic searches. After

189     duplicates' removal, we screened 110 records for relevance. There were 67 papers excluded

190     on a basis of evaluation of the title and abstract, leaving 43 articles to be assessed for

191     eligibility. Of these, 30 articles were included in the quantitative analysis. A PRISMA flow

192     diagram is provided in Figure 1. There were 9650 participants included in this systematic

193     review and meta-analysis.

194

**Study characteristics**

195

Table 1 depicts the characteristics of the included studies. There were six cohort studies[21, 34-

196

38], whilst the remaining studies were case - control[20, 22, 30, 39-59]. The definitions of

197

periodontitis varied among these studies, while the definition of preeclampsia is presented

198

comparatively consistent as blood pressure $\geq$ 140/90 mmHg and proteinuria during second

199

trimester of gestation. The oral examination was conducted at different timepoints among

200

studies, during pregnancy[21, 34-39, 41, 47, 55, 56, 59], within 24 to 48 hours prior to delivery[40, 43, 52, 54]

201

or after delivery[22, 30, 37, 42, 44-46, 48-51, 53, 57, 58] The sample size ranged from 40 participants [20] to

202

1240 subjects [53]. There were seven studies with no evidence of an association between

203

periodontitis and preeclampsia[22, 30, 35, 39, 45, 46, 50, 55, 56], while the remaining studies reported a

204

positive association.

205

206

**Risk of bias of included studies**

207

Newcastle Ottawa Scale was used to evaluate the quality of evidence of these reports. Two

208

reviewers marked the scores for each paper based on the tool provided by the Scale. Nine

209

studies [21, 37, 38, 48, 51, 54-57] obtained the maximum score in Selection outcome while fourteen

210

studies were marked with maximum score in the Comparability outcome and none of the

211

studies could achieve ultimately 3 marks in the Exposure outcome. Table 2 describes the

212

evaluation of risk of bias for this review.

213

214

**Synthesis of results**

215

The results of the meta-analysis showed that periodontitis was associated with increased risk

216

for preeclampsia (OR 3.18, 95% CI 2.26 – 4.48, p<0.00001; Figure 2). The heterogeneity

217

was high ($I^2$=81%, p<0.00001) revealing a significant variation among studies.

218

**Subgroup analysis.**

According to the study type, the results revealed the increased risk of preeclampsia in periodontitis patients in the cohort studies (OR 4.19, 95% CI 2.23 – 7.87, p < 0.00001; Figure 3) and in case-control studies (OR 2.96, 95% CI 2.00 – 4.39, p<0.00001; Figure 4). Heterogeneity was moderate for cohort ($I^2$ = 55%, p = 0.05) but high for case-control ($I^2$ = 83%, p < 0.00001).

When analyzing according to the national income, increased risk of preeclampsia were found in periodontitis group in high-income countries (OR 2.67, 95% CI 1.59 – 4.49, p = 0.0002; Figure 5), upper middle-income countries (OR 2.40, 95% CI 1.73 – 3.31, p < 0.00001; Figure 6) and especially lower-middle income countries (OR 6.7, 95% CI 2.61 – 17.19, p < 0.0001; Figure 7). Heterogeneity in the group of high-income, upper middle-income and lower middle-income countries were moderate ($I^2$ = 59%, p = 0.006; $I^2$ = 64%, p = 0.003; $I^2$ = 86%, p<0.00001, respectively).

When the results were analyzed according to the definition of periodontitis, an increased risk of preeclampsia was observed in all subgroups, including PD only (OR 3.13, 95% CI 1.51 – 6.50, p = 0.002), CAL and PD (OR 3.30, 95% CI 2.02 – 5.41, p<0.00001), CAL alone (OR 2.74, 95% CI 1.50 – 5.01, p = 0.001). Heterogeneity was moderate for the subgroups in which periodontitis was defined by PD alone ($I^2$ = 59%, p = 0.03) and CAL alone ($I^2$ = 66%, p = 0.01), while significantly high heterogeneity was found in the subgroup which periodontitis was defined by CAL and PD ( $I^2$ = 86%, p<0.00001).

When analyzing the periodontal condition between both groups, mean CAL was statistically higher in the preeclamptic patients than in the healthy group (MD = 0.62, 95% CI 0.27 – 0.98, p=0.0006). Likewise, the preeclamptic group had a statistically higher mean PD compared to healthy group (MD = 0.79, 95% CI -0.47 – 1.11, p<0.00001). The heterogeneity was 98% in both subgroup analysis, ($I^2$ = 98%, p<0.00001).

244

**Publication bias**

246 The funnel plot for the association between periodontitis and preeclampsia revealed the

247 symmetry (Figure 13). There was no publication bias.

248

**DISCUSSION:**

250

251 The aim of this meta-analysis was to re-evaluate the potential association between

252 preeclampsia and periodontitis. The results confirm that periodontitis is a risk factor for

253 preeclampsia, which was similar to the finding of a meta-analysis in 2013 by Sgolastra et al[18].

254 Our review has fifteen additional studies with three more cohort studies considerably

255 increasing the  sample size and  hence generating more robust effect sizes and significance

256 levels.

257

258 By stratifying according to study designs, periodontitis and preeclampsia showed significant

259 associations in both case-control and cohort studies, whereas Sgolastra et al[18] could not report

260 the statistical significance in the subgroup analysis of cohort studies (OR 2.2, 95% CI 0.66 –

261 7.36, p=0.2). As a review of cohort studies provides higher level of evidence compared to

262 case-control studies[60],   our data considerably strengthens the evidence of a positive

263 association between preeclampsia and periodontitis. Moreover, the heterogeneity in an

264 analysis of cohort studies in our study ($I^2$=55%, p=0.05) was substantially lower than in a

265 study by Sgolastra et al ($I^2$=89%, p=0.0001), again strengthening the reliability of our results.

266 [18]

267

268 When analyzed according to the definition of periodontitis, three subgroup analysis with

269 studies defining periodontitis by PD alone, CAL and PD and CAL alone showed statistically

270 significant differences, whereas the previous meta-analysis showed only significance with a

271 subgroup analyzing periodontitis by CAL and PD. This could be explained by the number of

272 studies included in our review was more than in the previous analysis, thus, providing a more

273 comprehensive finding. However, according to the most recent case definition developed by

274 the Centre for Disease Control and Prevention in partnership with the American Academy of

275 Periodontology, the diagnostic criteria of periodontitis is at least 2 interproximal sites with

276 the minimum of attachment loss of 3 mm and at least 2 interproximal sites with the minimum

277 pocket depth of 4 mm (not on the same tooth) or one site with pocket depth $\geq 5$ mm. [61]

278 Moreover, pregnant women who were preeclamptic had higher mean CAL and PD, however,

279 the heterogeneity in both subgroup analysis was high, indicating significant variations among

280 these studies in each subgroup. This could result from the difference in the periodontal probes

281 used in the dental examination in each study.

282 Several mechanisms have been proposed for the link between periodontitis and preeclampsia.

283 Higher levels of some periodontal pathogens such as *P.gingivalis* and *F. nucleatum* were

284 found in placenta of patients with preeclampsia. [62]Moreover, inflammatory responses

285 including the shifting of Th2 toward Th1, increasing oxidative stress, anti-angiogenic

286 proteins, vascular endothelial growth factor receptor 1 and complement C5a could potentially

287 enhance the development of preeclampsia. [63] Ananth et al. has reported the association

288 between intrauterine growth restriction and maternal periodontitis. [64]Since severe and early

289 onset preeclampsia were associated significantly with fetal growth restriction, this could

290 contribute to the mechanism underlying the association between preeclampsia and

291 periodontitis.[65] Furthermore, the mechanisms might be a reflection of dietary patterns.

292 Recently, some evidence has indicated that pathogenesis of preeclampsia involves maternal

293    gut microbiota, specifically, high-fiber diet which promote short chain fatty acid production

294    and are associated with reduced risk of preeclampsia.[66] Similarly, high-fiber foods such as

295    fruit and grains have been linked to the reduction of the progression of periodontal disease,

296    suggesting the role of dietary intake in the potential relationship between preeclampsia and

297    periodontal disease. [66, 67] However, future studies are required to elucidate these hypotheses.

298

299    When analyzed according to national income, this review revealed the significant difference

300    in the subgroup analysis of high-income and upper middle-income countries (OR=2.67 and

301    OR=2.40, respectively). Moreover, the subgroup analysis with lower middle-income

302    countries, which generated Odds ratio of 6.70, indicated the considerable significance in the

303    relationship between periodontitis and preeclampsia in this specific country group. Lower

304    middle-income countries have poorer oral health condition. [68] Moreover, they have lower

305    access to oral health care services (35%) to compare with upper middle-income (60-75%) and

306    high-income countries (82%).[69] Therefore, it is notably important to improve the access to

307    oral health services in pregnant women to lessen the risk of having preeclampsia.

308

309    We used Newcastle Ottawa Scale to evaluate the risk of bias and found twenty studies with

310    moderate risk of bias and ten remaining studies with high risk of bias. Furthermore, there was

311    no publication bias.

312

313    The strength of this systematic review and meta-analysis includes the large sample size of

314    9650 subjects. Six cohort studies comprising 2840 subjects were analyzed and revealed the

315    statistically significant difference. Because cohort approach is the best methodical design in

316    epidemiology, this finding provided the reliable evidence for the association between

317    periodontitis and preeclampsia. Furthermore, by stratifying into subgroup analysis of national income, our review has pointed out the association between these two diseases differed according to economical inequalities, thus, providing recommendation for health policy improvement. Pregnant women in low socioeconomical areas should be given access to oral healthcare services and encouraged to have their periodontal health checked and treated during pregnancy to potentially lower the risk of preeclampsia as well as other pregnancy complications. Jeffcoat et al reported non-surgical periodontal therapy could significantly reduce the medical costs for pregnant women by 73.7%. [70] There were also some limitations. Firstly, the heterogeneity of the overall analysis for the association between periodontitis and preeclampsia was high, pointing out the variations among studies included. This could be due to the synthesis of cohort and case-control studies in our review. Secondly, the general consensus in the definition and diagnosis of periodontitis was not clear enough which could influence the results of our meta-analysis. Therefore, future studies could take these into consideration and confirm our results.

**CONCLUSIONS AND IMPLICATIONS**

This meta-analysis not only confirms previous findings of an association between periodontitis and preeclampsia but also shows larger effect size overall and specifically for lower-middle income countries in comparison to high and upper-middle income countries. Our results warrant future studies to investigate the mechanisms of this association and whether targeted interventions to prevent or treat periodontitis preconception or during pregnancy can lead to better pregnancy outcomes.

**FUNDING**

This research did not receive any funding support from any parts.

343 **REFERENCE**

344 1.      Duley L. The global impact of pre-eclampsia and eclampsia. Semin Perinatol.
345 2009;33(3):130-7.
346 2.      Hung T-H, Hsieh Ts-Ta, Chen S-F. Risk of abnormal fetal growth in women with early-
347 and late-onset preeclampsia. Pregnancy hypertension. 2018;12:201-6.
348 3.      Turbeville HR, Sasser JM. Preeclampsia beyond pregnancy: long-term consequences for
349 mother and child. American journal of physiology Renal physiology. 2020;318(6):F1315-F26.
350 4.      Stevens W, Shih T, Incerti D, Ton TGN, Lee HC, Peneva D, et al. Short-term costs of
351 preeclampsia to the United States health care system. American Journal of Obstetrics &
352 Gynecology. 2017;217(3):237-48.e16.
353 5.      Gesase N, Miranda-Rius J, Brunet-Llobet L, Lahor-Soler E, Mahande MJ, Masenga G. The
354 association between periodontal disease and adverse pregnancy outcomes in Northern
355 Tanzania: a cross-sectional study. African health sciences. 2018;18(3):601-11.
356 6.      Govindasamy R, Dhanasekaran M, Varghese S, Balaji V, Karthikeyan B, Christopher A.
357 Maternal risk factors and periodontal disease: A cross-sectional study among postpartum
358 mothers in Tamil Nadu. Journal of Pharmacy And Bioallied Sciences. 2017;9(5):50-4.
359 7.      Alchalabi HA, Al Habashneh R, Jabali OA, Khader YS. Association between periodontal
360 disease and adverse pregnancy outcomes in a cohort of pregnant women in Jordan. Clin Exp
361 Obstet Gynecol. 2013;40(3):399-402.
362 8.      Piscoya MD, Ximenes RA, Silva GM, Jamelli SR, Coutinho SB. Periodontitis-associated risk
363 factors in pregnant women. Clinics (Sao Paulo, Brazil). 2012;67(1):27-33.
364 9.      Bui FQ, Almeida-da-Silva CLC, Huynh B, Trinh A, Liu J, Woodward J, et al. Association
365 between periodontal pathogens and systemic disease. Biomedical Journal. 2019;42(1):27-35.
366 10.     Winning L, Linden GJ. Periodontitis and systemic disease. BDJ Team. 2015;2(10):15163.
367 11.     Corbella S, Taschieri S, Francetti L, De Siena F, Del Fabbro M. Periodontal disease as a
368 risk factor for adverse pregnancy outcomes: a systematic review and meta-analysis of case-
369 control studies. Odontology. 2012;100(2):232–40.
370 12.     Abariga SA, Whitcomb BW. Periodontitis and gestational diabetes mellitus: a systematic
371 review and meta-analysis of observational studies. BMC Pregnancy Childbirth 2016;16(344).
372 13.     Kivimäki M, Batty GD, Pentti J, Shipley MJ, Sipilä PN, Nyberg ST, et al. Association
373 between socioeconomic status and the development of mental and physical health conditions in
374 adulthood: a multi-cohort study. The Lancet Public Health. 2020;5(3):e140-e9.
375 14.     Bo S, Menato G, Bardelli C, Lezo A, Signorile A, Repetti E, et al. Low socioeconomic status
376 as a risk factor for gestational diabetes. Diabetes Metab. 2002;28(2):139-40.
377 15.     Silva LM, Coolman M, Steegers EAP, Jaddoe VWV, Moll HA, Hofman A, et al. Low
378 socioeconomic status is a risk factor for preeclampsia: the Generation R Study. Journal of
379 Hypertension. 2008;26(6).
380 16.     Peacock JL, Bland JM, Anderson HR. Preterm delivery: effects of socioeconomic factors,
381 psychological stress, smoking, alcohol, and caffeine. Bmj. 1995;311(7004):531-5.
382 17.     Ross KM, Dunkel Schetter C, McLemore MR, Chambers BD, Paynter RA, Baer R, et al.
383 Socioeconomic Status, Preeclampsia Risk and Gestational Length in Black and White Women. J
384 Racial Ethn Health Disparities. 2019;6(6):1182-91.
385 18.     Sgolastra F, Petrucci A, Severino M, Gatto R, Monaco A. Relationship between
386 periodontitis and pre-eclampsia: a meta-analysis. PloS one. 2013;8(8):e71387.
387 19.     Wei BJ, Chen YJ, Yu L, Wu B. Periodontal disease and risk of preeclampsia: a meta-
388 analysis of observational studies. PLoS One. 2013;8(8):e70901.
389 20.     Varshney S, Gautam A. Poor periodontal health as a risk factor for development of pre-
390 eclampsia in pregnant women. Journal of Indian Society of Periodontology. 2014;18(3):321-5.
391 21.     Soucy-Giguère L, Tétu A, Gauthier S, Morand M, Chandad F, Giguère Y, et al. Periodontal
392 Disease and Adverse Pregnancy Outcomes: A Prospective Study in a Low-Risk Population.
393 Journal of obstetrics and gynaecology Canada : JOGC = Journal d'obstetrique et gynecologie du
394 Canada : JOGC. 2016;38(4):346-50.

22.     Lafaurie GI, Gómez LA, Montenegro DA, De Avila J, Tamayo MC, Lancheros MC, et al. Periodontal condition is associated with adverse perinatal outcomes and premature rupture membranes in low income pregnant women in Bogota, Colombia: a case control study. The journal of maternal-fetal & neonatal medicine : the official journal of the European Association of Perinatal Medicine, the Federation of Asia and Oceania Perinatal Societies, the International Society of Perinatal Obstetricians. 2018;33(1):16-23.

23.     Lavigne SE, Forrest JL. An umbrella review of systematic reviews of the evidence of a causal relationship between periodontal disease and adverse pregnancy outcomes: A position paper from the Canadian Dental Hygienists Association. Can J Dent Hyg. 2020;54(2):92-100.

24.     Kunnen A, van Doormaal JJ, Abbas F, Aarnoudse JG, van Pampus MG, Faas MM. Periodontal disease and pre-eclampsia: a systematic review. J Clin Periodontol. 2010;37(12):1075-87.

25.     Australian Institute of Health and Welfare 2010. Socioeconomic variation in periodontitis among Australian adults 2004–06. Research report series no. 50 . Cat. no. DEN 207. Canberra: AIHW.

26.     Moher D, Liberati A, Tetzlaff J, Altman DG. Preferred Reporting Items for Systematic Reviews and Meta-Analyses: The PRISMA Statement. J Clin Epidemiol. 2009;62:1006-12.

27.     Stroup DF, Berlin JA, Morton SC, Olkin I, Williamson GD, Rennie D, et al. Meta-analysis of observational studies in epidemiology: a proposal for reporting. Meta-analysis Of Observational Studies in Epidemiology (MOOSE) group. Jama. 2000;283(15):2008-12.

28.     Country Income Groups (World Bank Classification) [Internet]. 2011. Available from: http://data.worldbank.org/about/country-classifications/country-and-lending-groups.

29.     Lo CK, Mertz D, Loeb M. Newcastle-Ottawa Scale: comparing reviewers' to authors' assessments. BMC Med Res Methodol. 2014;14(45).

30.     Taghzouti N, Xiong X, Gornitsky M, Chandad F, Voyer R, Gagnon G, et al. Periodontal Disease is Not Associated With Preeclampsia in Canadian Pregnant Women. Journal of periodontology. 2012;83(7):871-7.

31.     Higgins JP, Thompson SG. Quantifying heterogeneity in a metaanalysis. . Stat Med. 2002;21:1539-58.

32.     Egger M, Davey SG, Schneider M, Minder C. Bias in metaanalysis detected by a simple, graphical test. Br Med J. 1997;315:629-34.

33.     Orwin R. A fail-safe N for effect size in meta-analysis. . J Educ Stat. 1983;8:157-9.

34.     Boggess KA, Lieff S, Murtha AP, Moss K, Beck J, Offenbacher S. Maternal periodontal disease is associated with an increased risk for preeclampsia. Obstetrics and gynecology. 2003;101(2):227-31.

35.     Horton AL, Boggess KA, Moss KL, Beck J, Offenbacher S. Periodontal disease, oxidative stress, and risk for preeclampsia. Journal of periodontology. 2010;81(2):199-204.

36.     Kumar A, Basra M, Begum N, Rani V, Prasad S, Lamba AK, et al. Association of maternal periodontal health with adverse pregnancy outcome. The journal of obstetrics and gynaecology research. 2013;39(1):40-5.

37.     Ha JE, Jun JK, Ko HJ, Paik DI, Bae KH. Association between periodontitis and preeclampsia in never-smokers: a prospective study. Journal of clinical periodontology. 2014;41(9):869-74.

38.     Lee HJ, Ha JE, Bae KH. Synergistic effect of maternal obesity and periodontitis on preterm birth in women with pre-eclampsia: a prospective study. Journal of clinical periodontology. 2016;43(8):646-51.

39.     Chaparro A, Sanz A, Quintero A, Inostroza C, Ramirez V, Carrion F, et al. Increased inflammatory biomarkers in early pregnancy is associated with the development of pre-eclampsia in patients with periodontitis: a case control study. Journal of periodontal research. 2013;48(3):302-7.

40.     Canakci V, Canakci CF, Canakci H, Canakci E, Cicek Y, Ingec M, et al. Periodontal disease as a risk factor for pre-eclampsia: a case control study. The Australian & New Zealand journal of obstetrics & gynaecology. 2004;44(6):568-73.

41. Contreras A, Herrera JA, Soto JE, Arce RM, Jaramillo A, Botero JE. Periodontitis Is Associated With Preeclampsia in Pregnant Women. Journal of Periodontology. 2006;77(2).

42. Cota LO, Guimarães AN, Costa JE, Lorentz TC, Costa FO. Association between maternal periodontitis and an increased risk of preeclampsia. Journal of periodontology. 2006;77(12):2063-9.

43. Canakci V, Canakci CF, Yildirim A, Ingec M, Eltas A, Erturk A. Periodontal disease increases the risk of severe pre-eclampsia among pregnant women. J Clin Periodontol. 2007;34:639-45.

44. Siqueira FM, Cota LO, Costa JE, Haddad JP, Lana AM, Costa FO. Maternal periodontitis as a potential risk variable for preeclampsia: a case-control study. Journal of periodontology. 2008;79(2):207-15.

45. Lohsoonthorn V, Kungsadalpipob K, Chanchareonsook P, Limpongsanurak S, Vanichjakvong O, Sutdhibhisal S, et al. Maternal periodontal disease and risk of preeclampsia: a case-control study. American journal of hypertension. 2009;22(4):457-63.

46. Shetty M, Shetty PK, Ramesh A, Thomas B, Prabhu S, Rao A. Periodontal disease in pregnancy is a risk factor for preeclampsia. Acta obstetricia et gynecologica Scandinavica. 2009;89(5):718-21.

47. Politano GT, Passini R, Nomura ML, Velloso L, Morari J, Couto E. Correlation between periodontal disease, inflammatory alterations and pre-eclampsia. Journal of Dental Research. 2011;46:505-11.

48. Ha JE, Oh KJ, Yang HJ, Jun JK, Jin BH, Paik DI, et al. Oral Health Behaviors, Periodontal Disease, and Pathogens in Preeclampsia: A Case-Control Study in Korea. Journal of periodontology. 2011;82(12):1685-92.

49. Sayar F, Sadat Hoseini M, Abbaspour S. Effect of Periodontal Disease on Preeclampsia. Iranian J Publ Health. 2011;40(3):122-7.

50. Hirano E, Sugita N, Kikuchi A, Shimada Y, Sasahara J, Iwanaga R, et al. The association of Aggregatibacter actinomycetemcomitans with preeclampsia in a subset of Japanese pregnant women. J Clin Periodontol. 2012;39:229-38.

51. Moura da Silva G, Coutinho SB, Piscoya MD, Ximenes RA, Jamelli SR. Periodontitis as a Risk Factor for Preeclampsia. Journal of periodontology. 2012;83(11):1388-96.

52. Pralhad S, Thomas B, Kushtagi P. Periodontal Disease and Pregnancy Hypertension: A Clinical Correlation. Journal of periodontology. 2013;84(8):1118-25.

53. Desai K, Desai P, Duseja S, Kumar S, Mahendra J, Duseja S. Significance of maternal periodontal health in preeclampsia. Journal of International Society of Preventive & Community Dentistry. 2015;5(2):103-7.

54. Jaiman G, Nayak PA, Sharma S, Nagpal K. Maternal periodontal disease and preeclampsia in Jaipur population. Journal of Indian Society of Periodontology. 2018;22(1):50-4.

55. Pattanashetti JI, Nagathan VM, Rao SM. Evaluation of Periodontitis as a Risk for Preterm Birth among Preeclamptic and Non-Preeclamptic Pregnant Women - A Case Control Study. . Journal of clinical and diagnostic research : JCDR. 2013;7(8):1776–8.

56. Khalighinejad N, Aminoshariae A, Kulild JC, Mickel A. Apical Periodontitis, a Predictor Variable for Preeclampsia: A Case-control Study. Journal of endodontics. 2017;43(10):1611–4.

57. Khader YS, Jibreal M, Al-Omiri M, Amarin Z. Lack of association between periodontal parameters and preeclampsia. Journal of periodontology. 2006;77(10):1681–7.

58. Yaghini J, Mostajeran F, Afshari E, Naghsh N. Is periodontal disease related to preeclampsia? Dental research journal. 2012;9(6):770-3.

59. Kunnen A, Blaauw J, van Doormaal JJ, van Pampus MG, van der Schans CP, Aarnoudse JG, et al. Women with a recent history of early-onset pre-eclampsia have a worse periodontal condition. Journal of clinical periodontology. 2007;34(3):202-7.

60. Guyatt GH, Haynes RB, Jaeschke RZ, Cook DJ. Users' guides to the medical literature: XXV. evidence-based medicine: principles for applying the users' guides to patient care. Jama. 2000;284:1290-6.

61. Eke PI, Page RC, Wei L, Thornton-Evans G, Genco RJ. Update of the case definitions for population-based surveillance of periodontitis. J Periodontol. 2012;83(12):1449-54.

502    62.    Barak S, Oettinger-Barak O, Machtei EE, Sprecher H, Ohel G. Evidence of periopathogenic
503    microorganisms in placentas of women with preeclampsia. J Periodontol. 2007;78(4):670-6.
504    63.    Nourollahpour Shiadeh M, Behboodi Moghadam Z, Adam I, Saber V, Bagheri M, Rostami
505    A. Human infectious diseases and risk of preeclampsia: an updated review of the literature.
506    Infection. 2017;45(5):589-600.
507    64.    Ananth CV, Andrews HF, Papapanou PN, Ward AM, Bruzelius E, Conicella ML, et al.
508    History of periodontal treatment and risk for intrauterine growth restriction (IUGR). BMC Oral
509    Health. 2018;18(1):161.
510    65.    Odegård RA, Vatten LJ, Nilsen ST, Salvesen KA, Austgulen R. Preeclampsia and fetal
511    growth. Obstet Gynecol. 2000;96(6):950-5.
512    66.    Hu M, Eviston D, Hsu P, Mariño E, Chidgey A, Santner-Nanan B, et al. Decreased maternal
513    serum acetate and impaired fetal thymic and regulatory T cell development in preeclampsia.
514    Nature Communications. 2019;10(1):3031.
515    67.    Schwartz N, Kaye EK, Nunn ME, Spiro A, 3rd, Garcia RI. High-fiber foods reduce
516    periodontal disease progression in men aged 65 and older: the Veterans Affairs normative aging
517    study/Dental Longitudinal Study. Journal of the American Geriatrics Society. 2012;60(4):676-
518    83.
519    68.    Watt R, Sheiham A. Inequalities in oral health: a review of the evidence and
520    recommendations for action. Br Dent J 1999;187:6-12.
521    69.    Hosseinpoor AR, Itani L, Petersen PE. Socio-economic inequality in oral healthcare
522    coverage: results from the World Health Survey. J Dent Res. 2012;91(3):275-81.
523    70.    Jeffcoat MK, Jeffcoat RL, Gladowski PA, Bramson JB, Blum JJ. Impact of Periodontal
524    Therapy on General Health: Evidence from Insurance Data for Five Systemic Conditions.
525    American Journal of Preventive Medicine. 2014;47(2):166-74.
526

527

**Table 1. Descriptions of included studies**

| No | Reference | Country | Design | Participants and age | Definition of PE | Definition of PD | Examination Time | Finding (conclusion) | OR (95% CI) |
|---|---|---|---|---|---|---|---|---|---|
| 1 | Boggess et al. [34] (2003) | US | Cohort | 39 cases 763 controls | BP ≥ 140/90 on 2 separate occasions, and ≥ 1+ proteinuria on catheterized urine specimen | PD ≥ 4 and CAL ≥ 3mm without BOP. Mild: PD ≥ 4 mm or BOP on 1-15 teeth Severe: PD ≥ 4 mm on > 15 teeth Disease progression: ≥ 4 sites that increased ≥2mm in PD, resulting in ≥ 4 mm in PD | At the first or second prenatal visit and then repeated within 48 hours antepartum. Enrolled at < 26 weeks 'gestation and followed until delivery | Women were at higher risk for preeclampsia if they had severe periodontal disease at delivery or if they had periodontal disease progression during pregnancy | Severe periodontal disease: 2.4 (1.1- 5.3) Periodontal disease progression: 2.1 (1.0 – 4.4) |

| 2 | Canakci et al. [40] (2004) | Turkey | Case-control | 41 cases ,41 controls | BP ≥ 140/90 mmHg and proteinuria ≥ 300mg/24h or 2+ proteinuria on dip sticks, on 2 occasions ≥ 6h apart if 24h urine specimen is unavailable | ≥ 4 teeth with ≥ 1 sites with PD ≥ 4mm and BOP+ and CAL ≥ 3mm at the same site | within 48 h prior to delivery | multiple logistic regression results showed that pre-eclamptic patients were 3.47 times more likely to have periodontal disease than normotensive patients | 3.47 (1.07–11.95) |
| 3 | Contreras et al. [41] 2006 | Colombia | Case-control | 130 case, 243 controls | 2+ proteinuria, confirmed by ≥ 0.3 g proteinuria/24 hours and hypertension (≥ 140/ 90 mmHg) | ≥ 4 sites showed PD ≥4 mm), CAL ≥4 mm, and bleeding on probing. Incipient: CAL from 4 to 5 Moderate/severe: CAL ≥ 6 mm | between 26 to 36 weeks of pregnancy | Chronic periodontal disease was significantly associated with preeclampsia in pregnant women | 3.0 (1.91 - 4.87) |

| 4 | Cota et al. [42] 2006 | Brazil | Case-control | 109 cases, 479 controls | BP >140/90 mm Hg and ≥ 1+proteinuria after 20 weeks of gestation | ≥ 4 teeth with ≥ 1 sites with a PD ≥ 4 mm and CAL ≥ 3mm at the same site | Within 48 hours of delivery | Maternal periodontitis was determined to be associated with an increased risk of preeclampsia | 1.88 (1.1 - 3.0) |
|---|---|---|---|---|---|---|---|---|---|
| 5 | Khader et al.[57] 2006 | Jordan | Case-control | 115 cases 230 controls | Preeclampsia was defined as the development of blood pressure of ‡140/90 mmHg after 20 weeks of gestation, combined with proteinurea of at least 1+ on a midstream urine specimen or on a | Not mentioned | Within 24 hours after delivery | This study did not support the association between periodontal parameters and preeclampsia | |

| | | | | catheter specimen, provided urinary tract infection was not the contributing factor to the proteinuria in women who were known to be normotensive and nonproteinuric before pregnancy or in early pregnancy | | | | |
|---|---|---|---|---|---|---|---|---|
| 6 | Kunnen et al.[59] 2007 | Netherlands | Case-control | 17 cases, 35 controls | DBP ≥ 90mmHg on 2 occasions and proteinuria ≥ | PD ≥ 4mm | before 34 weeks of pregnancy | These results indicate that Caucasian | 7.9 (1.9–32.8) |

| | | | | | 30mg/dl (or 1+ on a urine dip stick) on ≥ 2 random specimens collected ≥4h apart. | Mild PD: BOP and PD ≥ 4mm on 1–15 sites<br><br>Severe: BOP and PD ≥ 4mm on > 15 sites | | women with a recent history of early-onset pre-eclampsia have a worse periodontal condition, as compared with women with uncomplicated deliveries | |
|---|---|---|---|---|---|---|---|---|---|
| 7 | Canakci et al.[43]<br>2007 | Turkey | Case-control | 20 Mild PE, 18 Severe PE, 21 Controls | DBP ≥ 90mmHg and proteinuria(300 mg/24h urine sample) and the presence of edema | Mild: BOP and ≥ 4mm PD on 1–15 sites<br>Severe: : BOP and ≥ 4mm PD on ≥15sites | within 48h preceding delivery | The results of multivariate logistic regression showed a highly significant association between mild to | Severe PD: 3.78 (1.77–12.74) |

| | | | | | Mild: BP ≥ 140/90mmHg on ≥ 2 occasions 6h apart, with or without proteinuria<br><br>Severe: SBP ≥ 160 or DBP ≥ 110mmHg on 2 occasions ≥6h apart and proteinuria ≥ 5g/24h urine sample or ≥ 3l on dip stick in ≥ 2 random clean-catch samples ≥ 4h apart | | | severe pre-eclampsia and severe periodontal disease | |

| 8 | Siqueira et al.[44] 2008 | Brazil | Case-control | 164 cases, 1042 controls | BP > 90mmHg on 2 occasions after 20 GW and ≥ 1+ proteinuria | ≥4mm and CAL ≥ 3mm at the same site in ≥ 4 teeth | within 48 hours of delivery | Maternal periodontitis is a risk factor associated with preeclampsia | 1.52 ( 1.01 - 2.29) |
|---|---|---|---|---|---|---|---|---|---|
| 9 | Lohsoonthorn et al.[45] 2009 | Thailand | Case-control | 150 cases, 150 controls | BP ≥ 140/90mmHg and proteinuria ≥ 30mg/dl (or 1+ on a urine dip stick) on ≥ 2 random specimens collected ≥ 4h apart. | Mild: ≥1 teeth with interproximal sites showing ≥4 mm CAL and ≥4 mm PD  Moderate: ≥2 nonadjacent teeth with interproximal sites showing ≥5 mm CAL and ≥4 mm PD  Severe: ≥2 nonadjacent teeth with interproximal sites showing ≥6 mm CAL and ≥4 mm PD | within 48 h after delivery | This study provides no convincing evidence that periodontal disease is associated with preeclampsia risk among Thai women | Severe PD: 0.92 ( 0.26–3.28) |

| 10 | Horton et al.[35] 2010 | US | Cohort | 34 cases (pree-clampsia) 757 controls (non pree-clampsia) | BP >140/90 mmHg and ≥ 1+ proteinuria on a catheterized urine specimen | Mild: <15 sites with the presence of one or more pockets ≥4 mm or one or more pockets with bleeding. Moderate/severe: ≥15 sites demonstrated a probing depth ≥4 mm | <26 weeks of gestation | Among women with periodontal disease, the presence of 8-isoprostane ≥ 75th percentile did not significantly increase the odds for the development of preeclampsia | 2.08 (0.65 - 6.60) |
| 11 | Shetty et al.[46] 2009 | India | Case-control | 30 cases 100 controls | BP > 140/90 mmHg on more than 2 occasions 4 hours apart and 1+ or more proteinuria by Dipstick on a | CAL of ≥ 3 mm and a PD of ≥ 4 mm. (The teeth examined were 16, 22, 24, 36, 42, and 44) | within 48 hours of delivery | periodontitis both at enrolment as well as within 48 hours of delivery may be associated with | Enrolment:5.78 (2.41–13.89) Delivery: 20.15 (4.55–89.29) |

| | | | | | random urine sample | | | an increased risk of preeclampsia. | |
|---|---|---|---|---|---|---|---|---|---|
| 12 | Politano et al.[47] 2011 | Brazil | Case-control | 58 cases 58 controls | increase in systolic arterial pressure (≥ 140 mmHg) and/or diastolic pressure (≥ 90mmHg) and proteinuria (≥ 300 mg/24 h), after 20 wk of gestation | two or more sites showed pocket formation (≥ 4 mm), clinical attachment level (≥ 4 mm) and bleeding on probing | after 20 wk of gestation | periodontal disease may increase the risk of pre-eclampsia | 3.73 (1.32–10.58) |
| 13 | Ha et al. [48] 2011 | Korea | Case-control | 16 cases 48 controls | BP >140/90 mm Hg on two separate occasions and ≥ 1+ proteinuria on | Localized periodontitis: periodontal clinical attachment loss ≥ 3.5 mm on two or three sites not on the same tooth. | 5 days after delivery | preeclampsia could be associated with the maternal periodontal condition and | Localized periodontitis: 4.79 (1.02 - 29.72) |

| | | | | | a random sample of urine | Generalized periodontitis: CAL ≥ 3.5 mm on ≥ 4 sites not on the same tooth | | interdental cleaning | Generalized periodontitis: 6.60 (1.25 - 41.61) |
|---|---|---|---|---|---|---|---|---|---|
| 14 | Sayar et al.[49] 2011 | Iran | Case-control | 105 cases 105 controls | blood pressure ≥140/90mmHg and proteinuria +1 | No definition specified (results based on periodontal parameters)<br><br>Mild: CAL ≤ 2mm<br><br>Moderate to Severe: CAL ≥ 3 mm | 48 hours after child delivery. | Preeclamptic cases significantly had higher attachment loss and gingival recession than the control group | 4.1 (1.5-11.5) |
| 15 | Taghzouti et al. [30] 2012 | Canada | Case-control | 92 cases 245 controls | BP ≥140/90 mm Hg on two occasions ≥4 hours apart after 20 weeks of | periodontitis is defined as ≥4 sites exhibiting PD ≥5 mm and CAL ≥3 mm at the same sites. | within 48 hours after delivery | This study does not support the hypothesis of an association between | 1.13 (0.59 - 2.17) |

| | | | | | | | | |
|---|---|---|---|---|---|---|---|---|
| | | | | gestation, and 0.3 g proteinuria on a 24-hour urine collection, or ≥1 on a dipstick | | | periodontal disease and preeclampsia | |
| 16 | Chaparro et al.[39] 2013 | Chile | Case-control | 11 cases 43 controls | During the second and third trimester of pregnancy, BP > 140/90 and proteinuria, which was considered to be present when one 24-h urine collection showed a total protein excretion ≥ 300 mg. | PD ≥ 4 mm and CAL ≥ 3 mm at the same site of ≥ 4 teeth, inflammation and bleeding on probing (BOP) | Blood samples were collected at enrolment  Gingival crevicular fluid samples were collected between 11–14 wk | Preeclamptic women shows increased levels of IL-6 in GCF and CRP in plasma during early pregnancy. Periodontal disease could contribute to systemic inflammation in early pregnancy via a local | 1.36 (0.252–7.372) |

| | | | | | | | increase of IL-6 and the systemic elevation of CRP. Therefore, both inflammatory markers could be involved in the relationship between periodontal disease and pre-eclampsia. | |
|---|---|---|---|---|---|---|---|---|
| 17 | Hirano et al.[50] 2012 | Japan | Case-control | 18 cases 109 controls | hypertension (systolic blood pressure > 140 mmHg and/or diastolic blood pressure > 90 | having over 60% of sites with CAL ≥ 3 mm | 5 days after labor | No statistically significant association between any of the periodontal clinical | 1.7 (1.1 - 2.7) |

| | | | | | mmHg) with proteinuria ( 300 mg/day) occurring after the 20th week of gestation, but being resolved by the 12th postpartum week | | | parameters or the presence of periodontitis and preeclampsia. | |
|---|---|---|---|---|---|---|---|---|---|
| 18 | Kumar et al.[36] 2012 | India | Cohort | 35 cases 305 controls | systolic blood pressure 140 mm of mercury and diastolic blood pressure 90 mm of mercury at two occasions at least 4 h apart after 20 weeks of gestation in a | clinical attachment loss and probing depth 4 ≥ mm in one or more sites were diagnosed as those with periodontitis | 14–20 weeks period of gestation | Maternal periodontitis is associated with an increased risk of pre-eclampsia | 7.48 (2.72–22.42) |

| | | | | | woman with previously normal blood pressure along with development of proteinuria | | | | |
|---|---|---|---|---|---|---|---|---|---|
| 19 | Da Silva et al.[15] 2012 | Brazil | Case-control | 284 cases 290 controls | a systolic blood pressure ≥ 140 mmHg or a diastolic pressure ≥90 mmHg and proteinuria ≥ 300 mg/24 hours or 2+ on dipsticks, developed after week 20 of gestation in previously | ≥4 teeth with ≥1 sites with a PD ≥4 mm and AL ≥3 mm in the same site | within 48 hours of childbirth | periodontitis was a risk factor for preeclampsia | 8.60 ( 3.92 - 18.88) |

| | | | | normotensive females | | | | |
|---|---|---|---|---|---|---|---|---|
| 20 | Pralhad et al.[52] 2012 | India | Case-control | 100 cases 100 controls | The resting blood pressure was ≥140/90 mmHg after 20 weeks of gestation with or without associated proteinuria | Any of the following is present: 1) OHI >3 69; 2) GI >1 (moderate-to-severe gingival inflammation); 3) mean PD >4 mm; and 4) CAL >3 mm | within 72 hours of their hospital admission for delivery | Periodontal disease is more prevalent in females with pregnancy hypertension | 5.5 (2.7 -11.4) |
| 21 | Yaghini et al.[58] 2012 | Iran | Case-control | 26 cases 25 controls | blood pressure >140/90 mmHg and > or = 1+ proteinuria on a catheterized urine specimen. | Not mentioned | 48 hours after delivery | Maternal periodontal disease during pregnancy is not associated with preeclampsia. | |
| 22 | Pattanashetti et al.[55] 2013 | India | Case-control | 100 cases 100 controls | Pregnancy induced hypertension | - Mild periodontal disease: One or more sites with probing | Sixth month of pregnancy and within 48 | Pregnant women with preeclampsia are | Periodontitis in the case group was 72%, in |

| | | | | | occurs after the 20th week of gestation and is characterized by: Hypertension – High blood pressure, usually higher than 140/90 mm Hg. The rise of blood pressure should be evident at least on two occasions, four or more hours apart. Edema – Demonstration of pitting oedema | depth ≥ 3 mm that bleed upon probing but less than 25 sites with probing depth ≥ 4mm. - Moderate/Severe: 15 or more sites with periodontal probing ≥4mm. - Worsening periodontal status was defined as four or more sites had increased by at least 2mm in pocket depth between the two oral health examinations | hours post-partum | at greater risk for preterm delivery if periodontal disease is present during pregnancy or progress during pregnancy and also the rate of preterm delivery is more in preeclamptic women having moderate to severe periodontal disease. | control group was 62%, p<0.001. |
|---|---|---|---|---|---|---|---|---|---|

| | | | | | over the ankles after 12 hours of bed rest or rapid gain in weight of more than 1 lb per week or more than 5 lb a month in the later month of pregnancy may be the easiest evidence of preeclampsia.<br><br>    Proteinuria – Presence of protein in 24 hours urine with more than 1gm per litre in 2 or more midstream | | | |
|---|---|---|---|---|---|---|---|---|

| | | | | | specimens obtained 6 hours apart in the absence of urinary tract infection is considered significant. | | | | |
|---|---|---|---|---|---|---|---|---|---|
| 23 | Ha et al.[37] 2014 | Korea | Cohort | 13 cases 270 controls | BP > 140/90 mmHg on two separate occasions, and at least 1+ proteinuria on a random urine screen after the 20th week of pregnancy | CAL ≥ 4.0 mm on two or more sites on different teeth. | 21–24 weeks of gestation | Periodontitis increased the risk of preeclampsia among never-smokers | 4.51 (1.13–17.96) |

| 24 | Varshney et al.[20] 2014 | India | Case-control | 20 cases 20 controls | BP ≥140/90 mm of Hg on two separate occasions after 20 week of gestation and ≥1+ proteinuria | PD ≥ 4 mm and CAL ≥ 3 mm at the same site on at least 4 different non-neighboring teeth | within the 48 h after delivery | Maternal clinical periodontal disease at delivery is associated with an increased risk for the development of pre-eclampsia, independent of the effects of maternal age, race, smoking, gestational age at delivery | 4.33 (1.15- 16.32) |
|----|----|----|----|----|----|----|----|----|----|
| 25 | Desai et al.[53] 2015 | India | Case-control | 120 cases 1120 controls | Blood pressure ≥140/90 mm Hg on two separate occasions after | PD ≥4 mm and CAL ≥3 mm at the same site in at least four teeth. | Periodontal examination was | Maternal clinical periodontal disease at delivery is | 19.898 (7.80-48.94) |

| | | | | | week 20 of gestation. | | performed 48h after delivery | associated with an increased risk for the development of pre-eclampsia, independent of the effects of maternal age, race, smoking, gestational age at delivery | |
|---|---|---|---|---|---|---|---|---|---|
| 26 | Soucy-Giguere et al.[21] 2015 | Canada | Cohort | 11 cases 237 controls | Not mentioned | the presence of at least one site with probing depths ≥ 4 mm and ≥ 10% bleeding on probing | | in the seven days following amniocentesis. | Pregnant women with periodontal disease were more likely to develop preeclampsia | RR 5.89 (1.24-28.05) |

| 27 | Lee et al.[38] 2016 | Korea | Cohort | 15 cases 313 controls | BP > 140/90 mmHg on two separate occasions with at least 1+ proteinuria on a random urine screen after the 20th week of pregnancy | two or more inter-proximal sites with CAL ≥4 mm that were not on the same tooth | at 21–24 weeks of gestation | The association was much stronger in women with both obesity and periodontitis | 15.94 (3.31–76.71) |
| 28 | Khalighinejad et al.[56] 2017 | USA | Case-control | 50 cases 50 controls | A systolic blood pressure ≥ 140 mm HG or a diastolic pressure ≥ 90 mm HG and proteinuria > 300 mg/24h developed after | The presence of 4 or more teeth with 1 or more sites with PD ≥4 mm and with clinical attachment loss ≥ 3mm at the same site | Before 26 weeks of gestation | Apical periodontitis was significantly more prevalent in the experimental group. | 2.23 (1.92-6.88) |

| | | | | | the 20th week of gestation | | | | |
|---|---|---|---|---|---|---|---|---|---|
| 29 | Lafaurie et al.[22] 2018 | Colombia | Case-control | 76 cases 304 controls | not mentioned | the patients were classified according to the presence of periodontal pockets (code 3: periodontal pockets of 4-5mm or code 4: periodontal pockets >5mm) | during the first week after birth | Periodontal pockets presence was not associated with preeclampsia. | 5.46 (1.84-16.1) |
| 30 | Jaiman et al.[54] 2018 | India | Case-control | 15 cases 15 controls | preeclampsia as the appearance of a diastolic blood pressure ≥90 mmHg mercury measured at two different occasions at least 4 h apart in combination with proteinuria (≥300 | According to the criteria of Löe and Silness. | 24 hours before delivery | The preeclamptic women were associated with significantly higher periodontitis and lower fetal birth weight than normotensive women | Periodontitis in case group was 93.3% and in control group was 33.3% (p<0.05) |

| | | | | | mg/24 h or +1 dipstick) developing after a gestational age of 20 weeks in a previously normotensive woman | | | | |
|---|---|---|---|---|---|---|---|---|---|
| BP: blood pressure; DBP: diastolic blood pressure; SBP: systolic blood pressure; GW: gestational week; PD: pocket depth; CAL: clinical attachment loss; OHI: oral hygiene index; GI: gingival index; BOP: bleeding on probing; GCF: gingival crevicular fluid | | | | | | | | | |

**Table 2. Risk of bias in included studies based on Newcastle-Ottawa scale.**

| Study, year | Selection (Max 4*) | Comparability (Max 2*) | Exposure (Max 3*) | Risk of bias |
|---|---|---|---|---|
| Boggess 2003 | *** | * | ** | High |
| Canakci 2004 | *** | ** | ** | Moderate |
| Contreras 2006 | *** | * | ** | High |
| Cota 2006 | *** | * | ** | High |
| Khader 2006 | **** | ** | ** | Moderate |
| Kunnen 2007 | *** | ** | ** | Moderate |
| Canakci 2007 | *** | ** | ** | Moderate |
| Siqueira 2008 | *** | * | ** | High |
| Lohsoonthorn 2009 | *** | ** | ** | Moderate |
| Horton 2010 | *** | * | ** | High |
| Shetty 2010 | *** | ** | ** | Moderate |
| Politano 2011 | *** | ** | ** | Moderate |
| Ha 2011 | **** | ** | ** | Moderate |
| Sayar 2011 | *** | ** | ** | Moderate |
| Taghzouti 2012 | *** | ** | ** | Moderate |
| Chaparro 2012 | *** | * | ** | High |
| Hirano 2012 | *** | * | ** | High |
| Kumar 2012 | *** | * | ** | High |
| da Silva 2012 | **** | ** | ** | Moderate |
| Pralhad 2012 | *** | ** | ** | Moderate |
| Yaghini 2012 | *** | ** | ** | Moderate |
| Pattanashetti 2013 | **** | ** | ** | Moderate |
| Ha 2014 | **** | ** | ** | Moderate |
| Varshney 2014 | *** | ** | ** | Moderate |
| Desai 2015 | *** | ** | ** | Moderate |
| Soucy-Giguere 2015 | **** | ** | *** | Low |
| Lee 2016 | **** | * | ** | Moderate |
| Khalighinejad 2017 | **** | ** | ** | Moderate |
| Lafaurie 2018 | *** | * | ** | High |
| Jaiman 2018 | **** | ** | ** | Moderate |

**FIGURE LEGENDS**

**Figure 1. Flow diagram of study selection**

**Figure 2. Forest plot for the association between periodontitis and preeclampsia.**

**Figure 3. Forest plot for the subgroup analysis according to the type of study design (cohort study).**

**Figure 4. Forest plot for the subgroup analysis according to the type of study design (case-control study)**

**Figure 5. Forest plot for the subgroup analysis according to the national income (high income countries)**

**Figure 6. Forest plot for the subgroup analysis according to the national income (upper middle-income countries)**

**Figure 7. Forest plot for the subgroup analysis according to the national income (lower middle- income countries)**

**Figure 8. Forest plot for the subgroup analysis according to the definition of periodontitis (PD alone)**

**Figure 9. Forest plot for the subgroup analysis according to the definition of periodontitis (PD and CAL)**

**Figure 10. Forest plot for the subgroup analysis according to the definition of periodontitis (CAL alone)**

**Figure 11. Forest plot for the subgroup analysis of mean CAL between preeclamptic and healthy groups.**

**Figure 12. Forest plot for the subgroup analysis of mean PD between preeclamptic and healthy groups.**

**Figure 13. Funnel plot for the association between periodontitis and preeclampsia.**

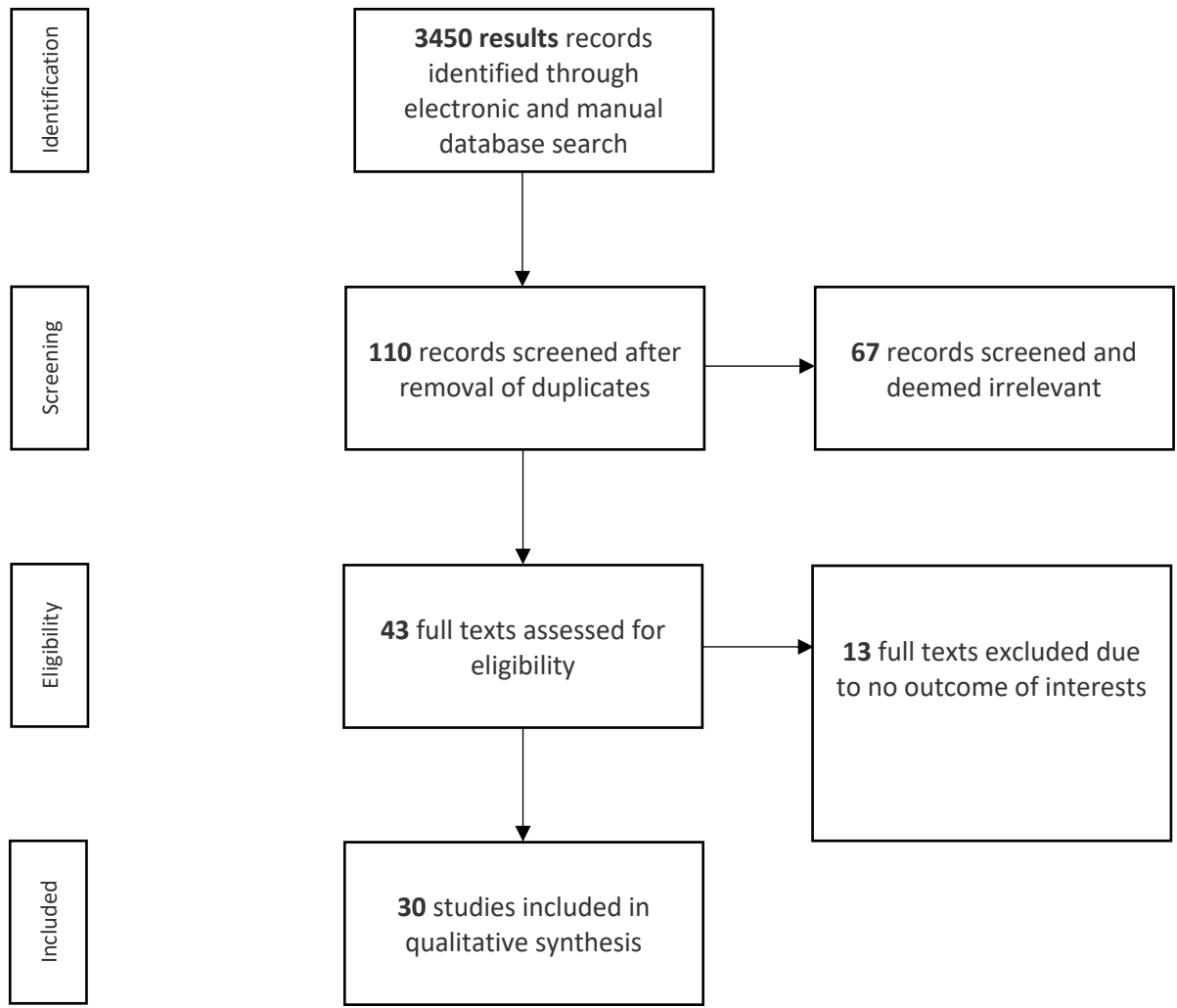

**Figure 1. Flow diagram of study selection**

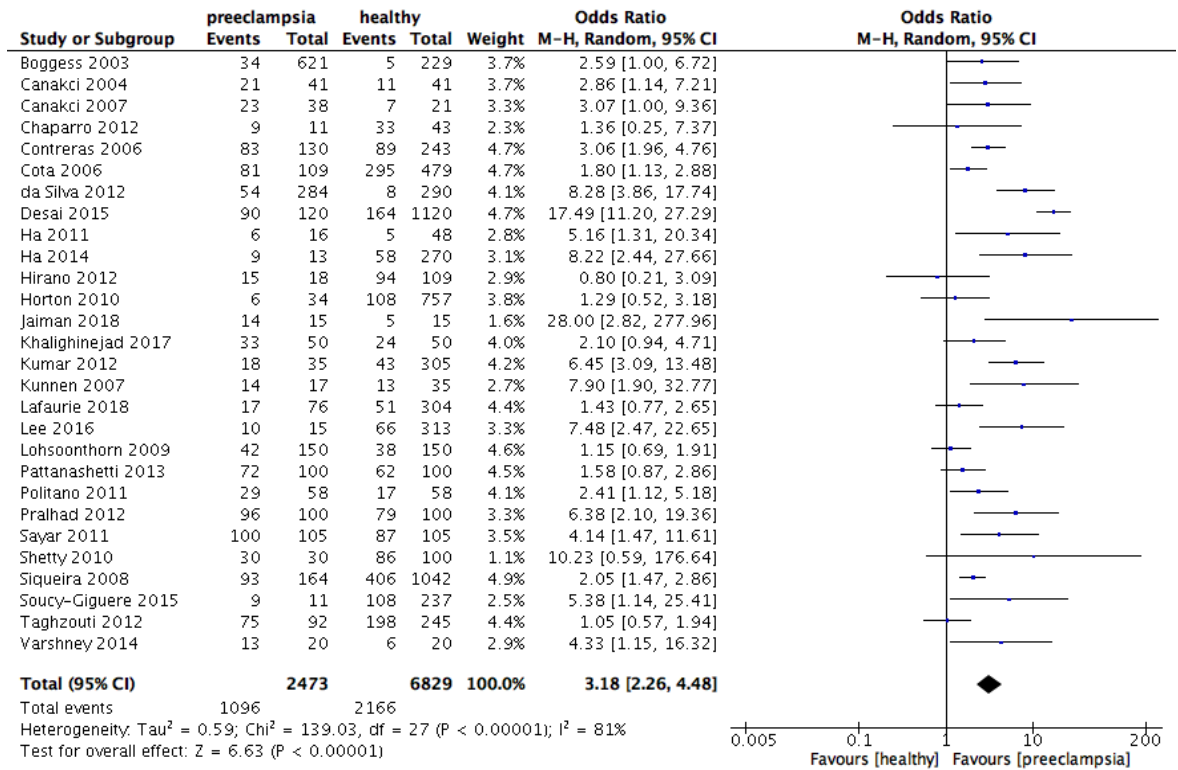

**Figure 2. Forest plot for the association between periodontitis and preeclampsia.**

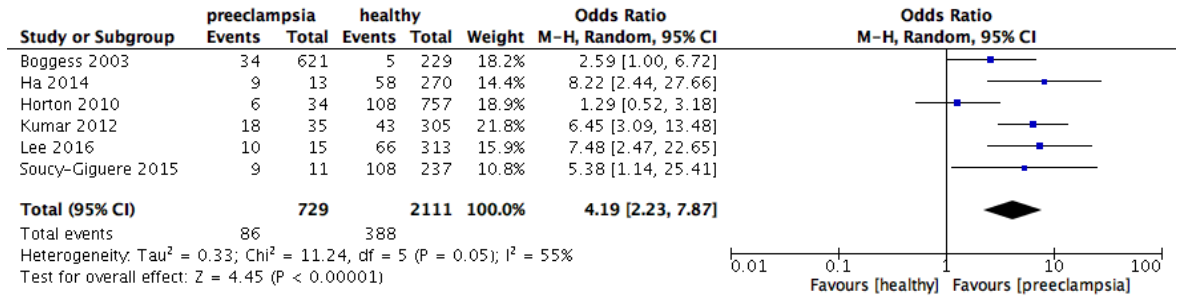

**Figure 3. Forest plot for the subgroup analysis according to the type of study design (cohort study)**

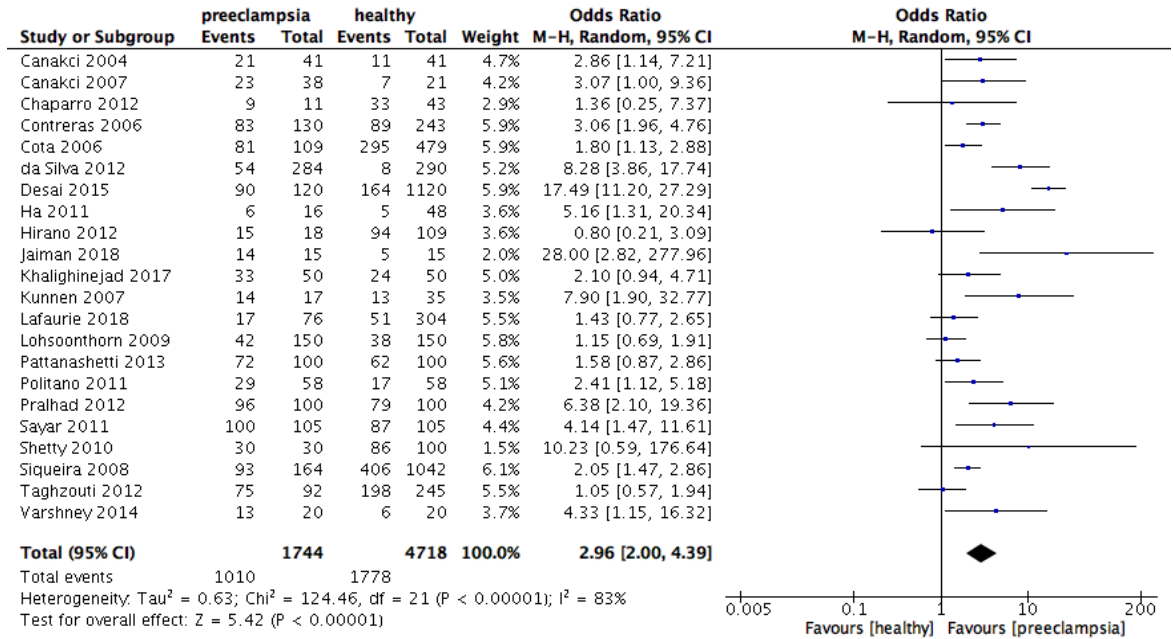

| Study or Subgroup | preeclampsia Events | Total | healthy Events | Total | Weight | Odds Ratio M-H, Random, 95% CI |
|---|---|---|---|---|---|---|
| Canakci 2004 | 21 | 41 | 11 | 41 | 4.7% | 2.86 [1.14, 7.21] |
| Canakci 2007 | 23 | 38 | 7 | 21 | 4.2% | 3.07 [1.00, 9.36] |
| Chaparro 2012 | 9 | 11 | 33 | 43 | 2.9% | 1.36 [0.25, 7.37] |
| Contreras 2006 | 83 | 130 | 89 | 243 | 5.9% | 3.06 [1.96, 4.76] |
| Cota 2006 | 81 | 109 | 295 | 479 | 5.9% | 1.80 [1.13, 2.88] |
| da Silva 2012 | 54 | 284 | 8 | 290 | 5.2% | 8.28 [3.86, 17.74] |
| Desai 2015 | 90 | 120 | 164 | 1120 | 5.9% | 17.49 [11.20, 27.29] |
| Ha 2011 | 6 | 16 | 5 | 48 | 3.6% | 5.16 [1.31, 20.34] |
| Hirano 2012 | 15 | 18 | 94 | 109 | 3.6% | 0.80 [0.21, 3.09] |
| Jaiman 2018 | 14 | 15 | 5 | 15 | 2.0% | 28.00 [2.82, 277.96] |
| Khalighinejad 2017 | 33 | 50 | 24 | 50 | 5.0% | 2.10 [0.94, 4.71] |
| Kunnen 2007 | 14 | 17 | 13 | 35 | 3.5% | 7.90 [1.90, 32.77] |
| Lafaurie 2018 | 17 | 76 | 51 | 304 | 5.5% | 1.43 [0.77, 2.65] |
| Lohsoonthorn 2009 | 42 | 150 | 38 | 150 | 5.8% | 1.15 [0.69, 1.91] |
| Pattanashetti 2013 | 72 | 100 | 62 | 100 | 5.6% | 1.58 [0.87, 2.86] |
| Politano 2011 | 29 | 58 | 17 | 58 | 5.1% | 2.41 [1.12, 5.18] |
| Pralhad 2012 | 96 | 100 | 79 | 100 | 4.2% | 6.38 [2.10, 19.36] |
| Sayar 2011 | 100 | 105 | 87 | 105 | 4.4% | 4.14 [1.47, 11.61] |
| Shetty 2010 | 30 | 30 | 86 | 100 | 1.5% | 10.23 [0.59, 176.64] |
| Siqueira 2008 | 93 | 164 | 406 | 1042 | 6.1% | 2.05 [1.47, 2.86] |
| Taghzouti 2012 | 75 | 92 | 198 | 245 | 5.5% | 1.05 [0.57, 1.94] |
| Varshney 2014 | 13 | 20 | 6 | 20 | 3.7% | 4.33 [1.15, 16.32] |
| | | | | | | |
| Total (95% CI) | | 1744 | | 4718 | 100.0% | 2.96 [2.00, 4.39] |
| Total events | 1010 | | 1778 | | | |

Heterogeneity: Tau² = 0.63; Chi² = 124.46, df = 21 (P < 0.00001); I² = 83%

Test for overall effect: Z = 5.42 (P < 0.00001)

**Figure 4. Forest plot for the subgroup analysis according to the type of study design (case-control study)**

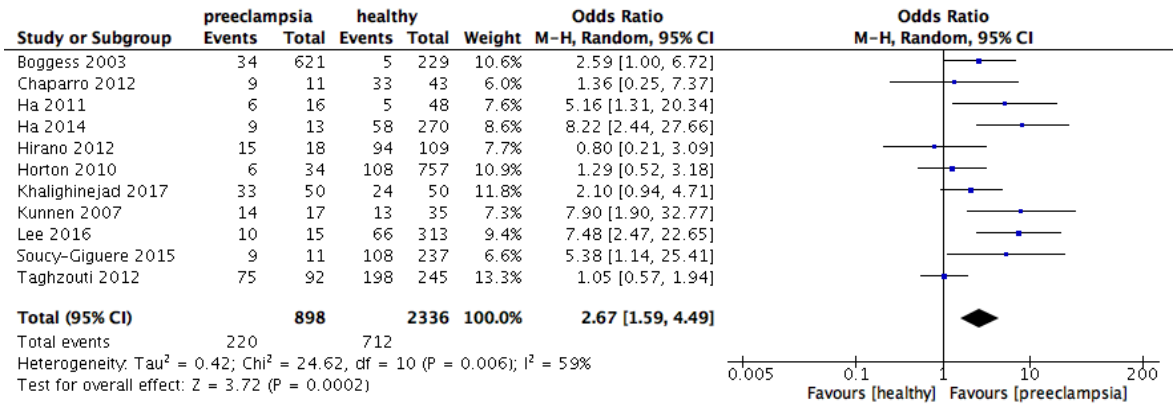

**Figure 5. Forest plot for the subgroup analysis according to the national income (high income countries)**

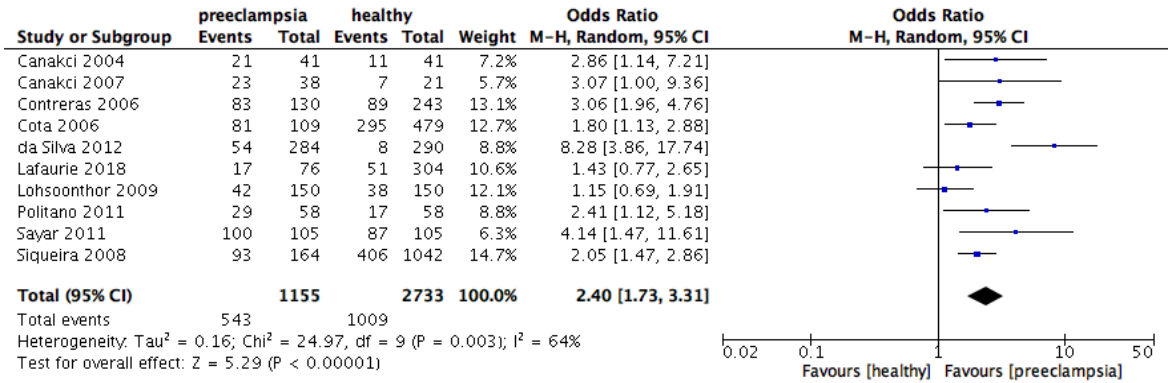

**Figure 6. Forest plot for the subgroup analysis according to the national income (upper middle-income countries)**

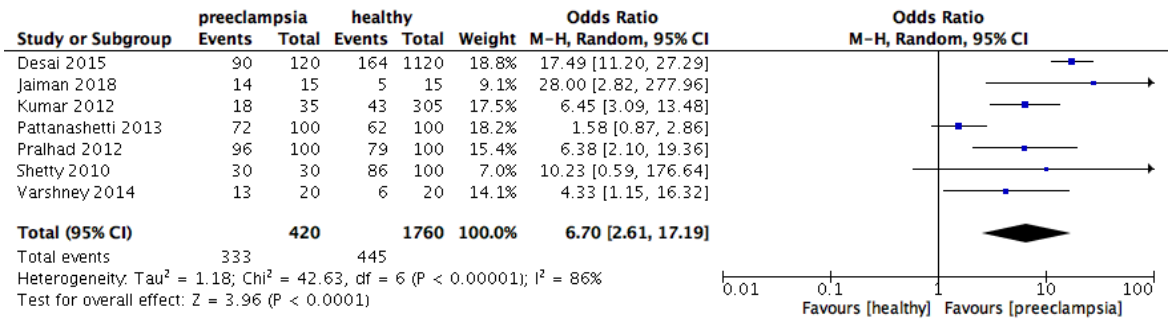

| Study or Subgroup | preeclampsia | | healthy | | | Odds Ratio | Odds Ratio |
| | Events | Total | Events | Total | Weight | M–H, Random, 95% CI | M–H, Random, 95% CI |
| --- | --- | --- | --- | --- | --- | --- | --- |
| Desai 2015 | 90 | 120 | 164 | 1120 | 18.8% | 17.49 [11.20, 27.29] | |
| Jaiman 2018 | 14 | 15 | 5 | 15 | 9.1% | 28.00 [2.82, 277.96] | |
| Kumar 2012 | 18 | 35 | 43 | 305 | 17.5% | 6.45 [3.09, 13.48] | |
| Pattanashetti 2013 | 72 | 100 | 62 | 100 | 18.2% | 1.58 [0.87, 2.86] | |
| Pralhad 2012 | 96 | 100 | 79 | 100 | 15.4% | 6.38 [2.10, 19.36] | |
| Shetty 2010 | 30 | 30 | 86 | 100 | 7.0% | 10.23 [0.59, 176.64] | |
| Varshney 2014 | 13 | 20 | 6 | 20 | 14.1% | 4.33 [1.15, 16.32] | |
| | | | | | | | |
| **Total (95% CI)** | | **420** | | **1760** | **100.0%** | **6.70 [2.61, 17.19]** | |
| Total events | 333 | | 445 | | | | |

Heterogeneity: Tau² = 1.18; Chi² = 42.63, df = 6 (P < 0.00001); I² = 86%
Test for overall effect: Z = 3.96 (P < 0.0001)

Favours [healthy]   Favours [preeclampsia]

**Figure 7. Forest plot for the subgroup analysis according to the national income (lower middle- income countries)**

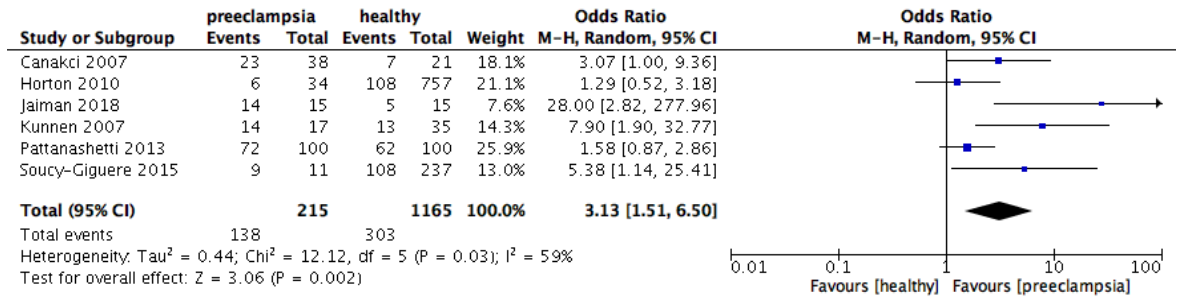

**Figure 8. Forest plot for the subgroup analysis according to the definition of periodontitis (PD alone)**

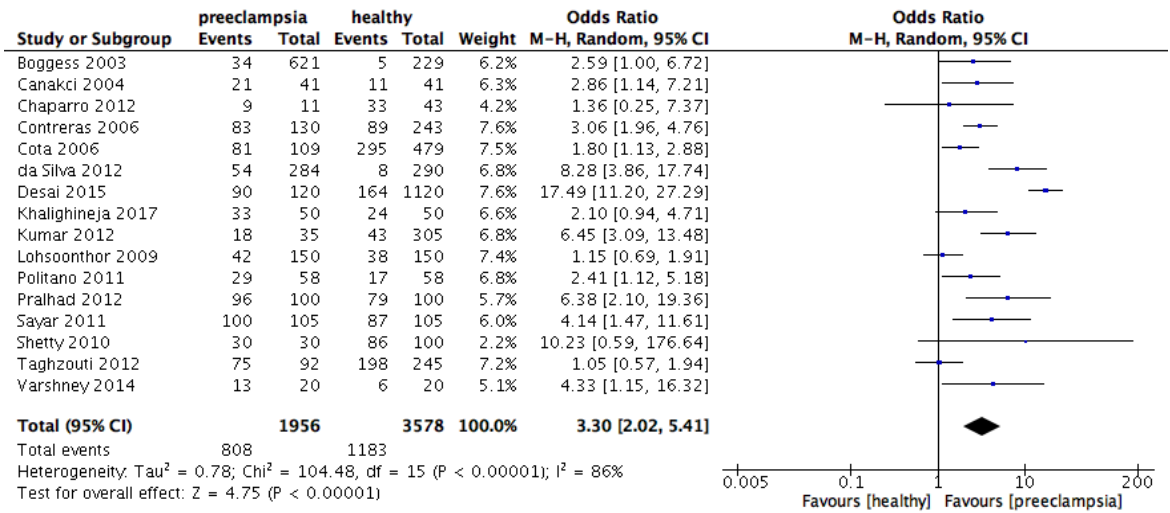

**Figure 9. Forest plot for the subgroup analysis according to the definition of periodontitis (PD and CAL)**

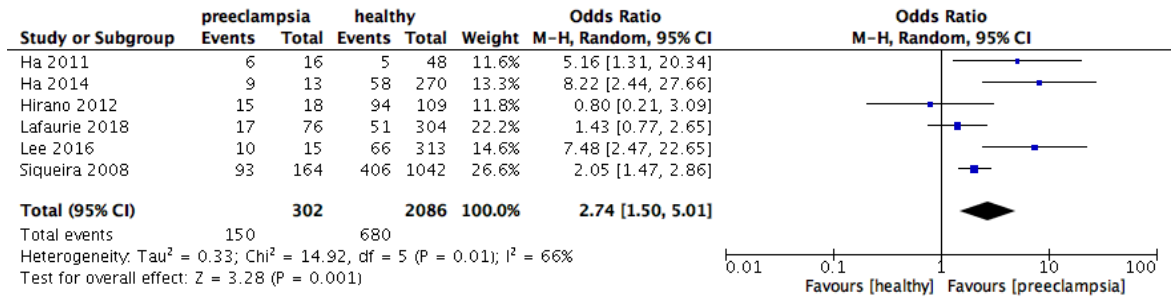

**Figure 10. Forest plot for the subgroup analysis according to the definition of periodontitis (CAL alone)**

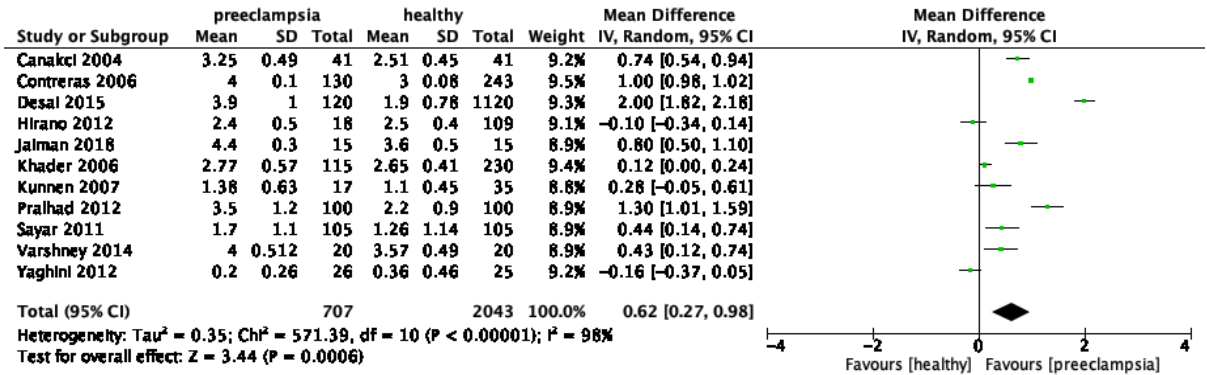

**Figure 11. Forest plot for the subgroup analysis of mean CAL between preeclamptic and healthy groups.**

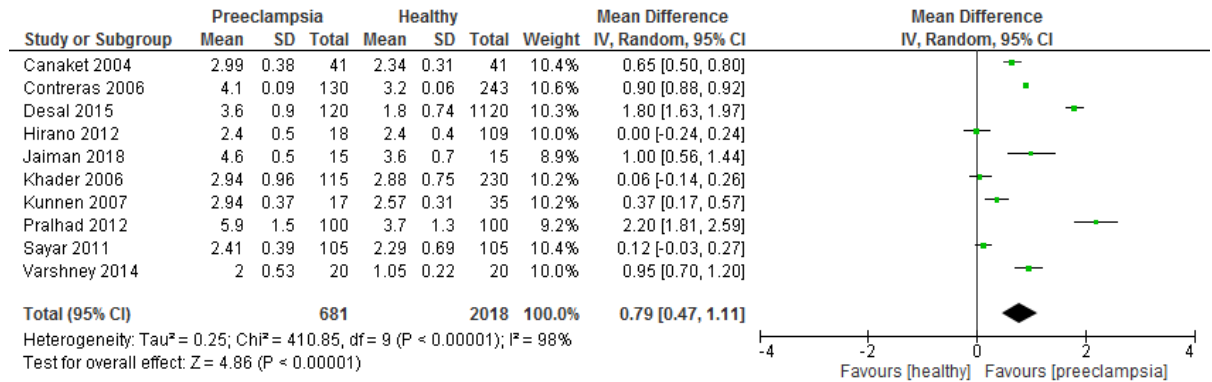

|  | Preeclampsia | | | Healthy | | | | Mean Difference | Mean Difference |
| Study or Subgroup | Mean | SD | Total | Mean | SD | Total | Weight | IV, Random, 95% CI | IV, Random, 95% CI |
|---|---|---|---|---|---|---|---|---|---|
| Canaket 2004 | 2.99 | 0.38 | 41 | 2.34 | 0.31 | 41 | 10.4% | 0.65 [0.50, 0.80] | |
| Contreras 2006 | 4.1 | 0.09 | 130 | 3.2 | 0.06 | 243 | 10.6% | 0.90 [0.88, 0.92] | |
| Desai 2015 | 3.6 | 0.9 | 120 | 1.8 | 0.74 | 1120 | 10.3% | 1.80 [1.63, 1.97] | |
| Hirano 2012 | 2.4 | 0.5 | 18 | 2.4 | 0.4 | 109 | 10.0% | 0.00 [-0.24, 0.24] | |
| Jaiman 2018 | 4.6 | 0.5 | 15 | 3.6 | 0.7 | 15 | 8.9% | 1.00 [0.56, 1.44] | |
| Khader 2006 | 2.94 | 0.96 | 115 | 2.88 | 0.75 | 230 | 10.2% | 0.06 [-0.14, 0.26] | |
| Kunnen 2007 | 2.94 | 0.37 | 17 | 2.57 | 0.31 | 35 | 10.2% | 0.37 [0.17, 0.57] | |
| Pralhad 2012 | 5.9 | 1.5 | 100 | 3.7 | 1.3 | 100 | 9.2% | 2.20 [1.81, 2.59] | |
| Sayar 2011 | 2.41 | 0.39 | 105 | 2.29 | 0.69 | 105 | 10.4% | 0.12 [-0.03, 0.27] | |
| Varshney 2014 | 2 | 0.53 | 20 | 1.05 | 0.22 | 20 | 10.0% | 0.95 [0.70, 1.20] | |
| **Total (95% CI)** | | | **681** | | | **2018** | **100.0%** | **0.79 [0.47, 1.11]** | |

Heterogeneity: Tau² = 0.25; Chi² = 410.85, df = 9 (P < 0.00001); I² = 98%
Test for overall effect: Z = 4.86 (P < 0.00001)

Favours [healthy]    Favours [preeclampsia]

**Figure 12. Forest plot for the subgroup analysis of mean PD between preeclamptic and healthy groups.**

ealthy groups.

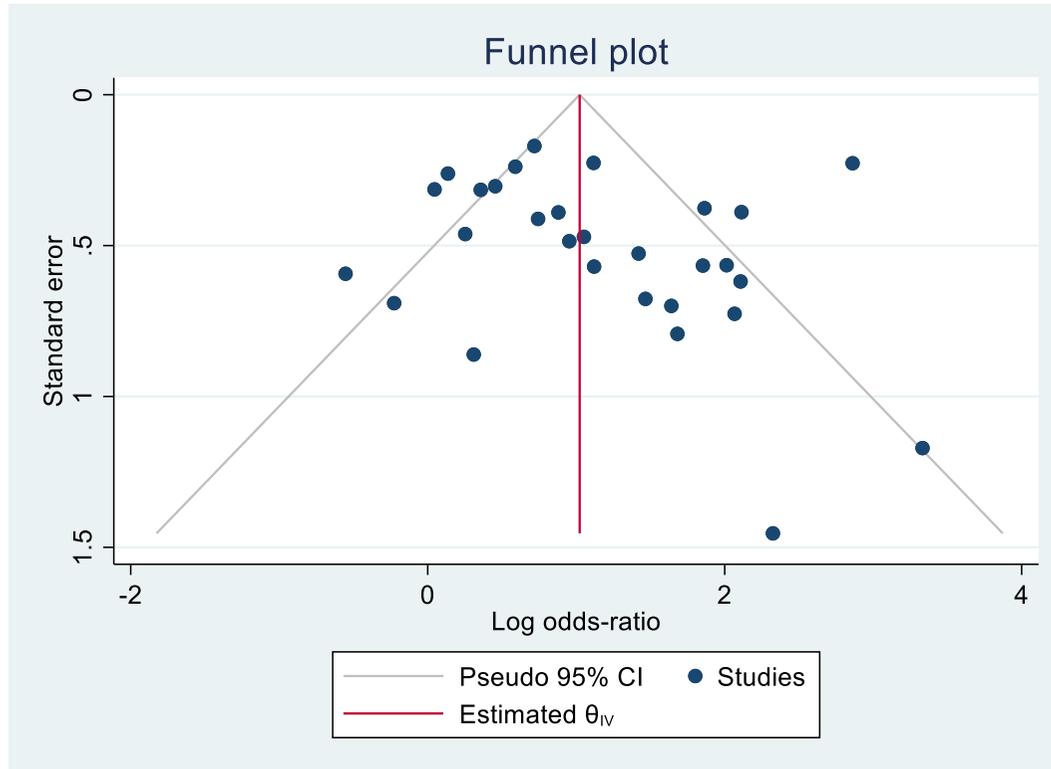

**Figure 13. Funnel plot for the association between periodontitis and preeclampsia.**